# Properties for liquid argon scintillation for dark matter searches

Akira Hitachi*

Research Institute for Science and Engineering, Waseda University, Tokyo 165-8555, Japan

A. Mozumder

Radiation Laboratory, University of Notre dame, Notre Dame, IN 46556-5674, USA

The scintillation yield for recoil Ar ions of 5 to 250 keV energy in liquid argon have been evaluated for direct dark matter searches. Lindhard theory is taken for estimating nuclear quenching. A theoretical model based on a biexcitonic diffusion-reaction mechanism is performed for electronic (scintillation) quenching. The electronic LET (linear energy transfer) is evaluated and used to obtain the initial track structure due to recoil Ar ions. The results are compared with experimental values reported for nuclear recoils from neutrons. The behavior of scintillation and ionization on the electric field are discussed.

(Dated: 14 March 2019)

## I. Introduction

Dark matter search is one of most important issue in physics and astrophysics today. The astronomical observations show the distribution of the unseen mass is different from that of ordinary matter [1]; the evidence suggests that the dark matter does not have efficient energy-loss mechanisms such as photon emission and inelastic collisions through molecular formation and condensation, etc. One of the most probable dark matter candidates is WIMP, Weakly Interacting Massive Particle [2-4]. Dual-phase noble-gas time projection chambers (TPC) are the most sensitive detectors for direct detection of galactic dark matter [5,6]. Liquid Ar is used for large scale WIMP detectors. Ionizing particles produces a prompt luminescence $S1$ and electrons in liquid phase. The electrons are extracted into gas phase where they produce proportional scintillation $S2$. Detection and selection of signals by WIMP, i.e. recoil Ar ions in a liquid argon dark matter detector, exploit the differences in scintillation efficiency, scintillation decay shape and charge collection due to electrons and recoil ions.

It is essential to know the scintillation efficiency for nuclear recoil produced by WIMPs striking nucleus in detector media to construct the recoil ion energy $E$. The energy of recoil ions is expected to be a few keV to a few hundreds keV. For such slow particles, only a part of energy $E_\eta$ given to electronic excitation can be used in ionization and scintillation detectors [7]. In addition to nuclear quenching, electronic quenching due to high-excitation density has to be considered in condensed phase scintillators [8]. It is the electronic LET (LET$_{el}$, linear energy transfer), not the electronic stopping power, that gives the number of excited species (excitons and electron-ion pairs) produced per unit length along the track [8,9]. The evaluation of LET$_{el}$ and the track structure are also needed for studying the recombination process with and without the external field.

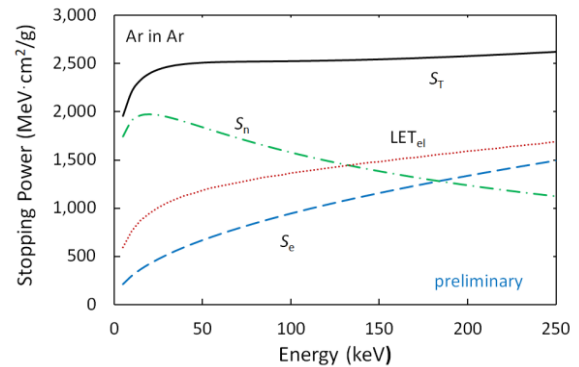

FIG. 1. Stopping powers and electronic LET for Ar ions in argon as a function energy.

In this paper, the so-called core and penumbra of heavy-ion track structure is considered and improved for understanding the track due to slow recoil ions. A quenching theory based on a biexcitonic diffusion-reaction mechanism is proposed and performed for electronic (scintillation) quenching [10]. The total quenching is obtained as a function of the energy and compared with the measurements. The sum signal, $S1$ and $S2$, is discussed for the reconstruction of recoil ion energy. We refer rare gases as argon and xenon, hereafter.

## II. Stopping power and electronic LET

For the interaction of slow ions with matter, the nuclear stopping power $S_n$ is of the same order of magnitude as the electronic stopping power $S_e$ [7,11] as shown in Fig. 1. We refer slow ions that the velocity $v$ is $v < v_0$, where $v_0 = e^2/\hbar \approx c/137 = 2.2\times10^8$ cm/sec is the Bohr velocity, $e$ is the charge of electron, and $c$ is the velocity of the light. The total stopping power $S_T$ is the sum of the two;

$$S_T = S_n + S_e. \quad (1)$$

The nuclear process follows the usual procedure of a screened Rutherford scattering. The Firsov potential is used

$$V(r) = \frac{Z_1 Z_2 e^2}{r} \Phi(r/a_{TFF}) \quad (2)$$

where $Z$ is the atomic number and suffix 1 and 2 are for projectile and the target, respectively. $\Phi$ is Fermi function. $a_{TFF}$ is the Thomas-Fermi-Firsov screening radius,

$$a_{TFF} = 0.8853 a_B / (Z_1^{1/2} + Z_2^{1/2})^{2/3}, \quad (3)$$

with the Bohr radius $a_B = \hbar/m_e e^2 = 0.529$ Å. Biersack gave the analytical expression for the nuclear stopping power [12]

$$S_n = \frac{2\pi a_{TFF} A_1 Z_1 Z_2 e^2}{A_1 + A_2} \frac{\ln \varepsilon}{\varepsilon(1-\varepsilon^C)}, \quad (4)$$

where $A$ is the atomic mass and $C = -1.49$. The energy $E$ in keV is converted to the dimensionless energy $\varepsilon$ by

$$\varepsilon = C_\varepsilon E = \frac{a_{TFF} A_2}{Z_1 Z_2 e^2 (A_1 + A_2)} E. \quad (5)$$

Eq. (5) becomes $\varepsilon = 11.5 Z_2^{-7/3} E$ for $Z_1 = Z_2$. The values of $C_\varepsilon$ is 0.01354 for Ar ions in argon.

Born's approximation is invalid for the low velocity region. Therefore, the Bethe theory of the stopping power is inapplicable to ion-atom collisions concerns here. The maximum energy given to the target atom is usually given by the kinematically limited maximum energy $Q_{max} = 4 E m_e / M$ where $m_e$ is the electron mass and $M$ is the mass of the projectile. For Ar ions, $4 m_e / M \sim 4/(2000 \times 40)$, then $Q_{max}$ is ~10 eV at $E = 200$ keV. The lowest excitation energy for Ar is about 11 eV. Then, the crude estimate shows no electronic excitation below 200 keV and $S_e$ becomes zero. However, no kinematic cut-off in electronic excitation have been observed. The kinematic limitation implies the ordinary approach in ion-atom collisions is not adequate. Lindhard et al. have taken a dielectric-response approximation [13]. The source density for incident heavy ion of charge $Z_1 e$ and velocity $\mathbf{v}$ is given by $\rho_0(\mathbf{r},t) = Z_1 e \delta(\mathbf{r} - \mathbf{v}t)$, corresponding to rectilinear motion. This charge causes polarization and changes the longitudinal dielectric constant in the electron gas. Consequently, the incident particle receives the electric force opposite direction. The stopping power $S$ is given by $S = Z_1 e \mathbf{E} \cdot (\mathbf{v}/v)$, where $\mathbf{E}$ is the electric field, for the electron gas. They applied the theory for ions-atom collisions based on the Thomas-Fermi treatment. For slow ions, the electronic stopping power is scale as $(d\varepsilon/d\rho)_e \approx k \varepsilon^{1/2}$, where $\rho$ is the dimensionless range and $k$ is the electronic stopping constant. The results expressed for a projectile ion and a target atom, to a first approximation [14],

$$S_e = \xi_e \times 8\pi e^2 a_B \frac{Z_1 Z_2}{(Z_1^{2/3} + Z_2^{2/3})^{3/2}} \frac{v}{v_0},$$

$$\text{with } \xi_e \approx Z_1^{1/6}, \quad (6)$$

where $k$ is expressed as $k = 0.133 Z_2^{2/3} A_2^{-1/2}$ for $Z_1 = Z_2$. For most cases, $k = 0.1 \sim 0.2$. We have $k = 0.145$ for Ar ions in argon. The stopping power cross sections discussed above give the same values as those in the HMI tables [15] at a low $E$. The stopping powers in liquid were obtained simply taking the density (1.40 g/cm$^3$ for liquid argon) into account, since uncertainty in the theory is larger than the difference in stopping powers in gas and liquid.

Lindhard et al. [7] have performed numerical calculations and given the partition of energy $\varepsilon$ of the incident ions to nuclear motion $\upsilon$ and electronic excitation $\eta$ in case of $Z_1 = Z_2$. Following expressions was taken for $k = 0.15$ from Fig. 3 in Ref. 7.

$$\eta = 0.427 \varepsilon^{1.193}, \quad \varepsilon < 4. \quad (7)$$

The nuclear quenching factor (Lindhard factor) $q_{nc}$ was defined as

$$q_{nc} = \eta / \varepsilon = E_\eta / E. \quad (8)$$

The electronic LET is given by [8,9],

$$LET_{el} = -\frac{dE_\eta}{dx} = -\frac{dE_\eta}{dE} \cdot \frac{dE}{dx} = \frac{d\eta}{d\varepsilon} \cdot S_T. \quad (9)$$

$LET_{el}$ is obtained analytically with Eqs. (1), (4), (6), (7) and (8). An averaged form is simply:

$$<LET_{el}> = -E_\eta / R = q_{nc} <LET> \quad (10)$$

where $R$ is the range of ions and $<LET>$ is the averaged LET. The stopping powers and the

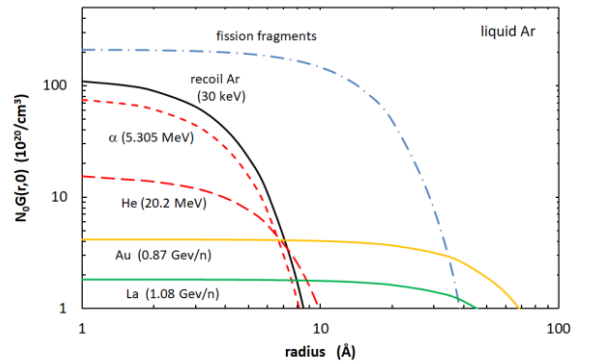

FIG. 2. The initial distribution of excited species in the track core produced by various particles in liquid argon. Solid curve shows 30 keV recoil Ar ions. La and Au ions are relativistic.

electronic LET are shown as a function of energy in Fig. 1

## III. Heavy ion track structure

The heavy-ion track can be regarded as a co-axial cylindrical geometry consist of the high-density core and surrounding less dense penumbra [16]. The core is mostly due to glancing collisions and the penumbra is formed by δ-rays produced by knock-on collisions. The local dose distribution in penumbra is roughly scale as $r^{-2}$, where $r$ is the radius, and the diameter is given by the range of maximum energy δ-rays. Since electronic quenching take place in the core of heavy ions, we consider excitation density in the core. The total (electronic) energy $T$ given to the liquid is divided into the core $T_c$ and into the penumbra $T_p$. The energy $T_s$ available for scintillation is,

$$T_s = qT = q_c T_c + T_p \qquad (11)$$

where $q$ and $q_c$ are the overall quenching factor and that in the core, respectively. The initial radial dose distribution in track core may be approximated as Gaussian with the core radius $a_0$ [10],

$$D_c = \frac{LET_c}{\pi a_0^2} \exp(-r^2 / a_0^2) \;, \qquad (12)$$

where $LET_c$ is the linear energy transfer in the core and is the sum of contribution of glancing collision $LET/2$ and $LET_c^\delta$ left by δ-rays. The core radius $a_0$ is given by Bohr's impulse principle, $r_B = \hbar v/2E_1$, where $\hbar$ is Plank's constant divided by $2\pi$, $v$ is the velocity of incident ion and $E_1$ is the energy of lowest electronic excited state of the medium. For liquid argon, $E_1 = 12.1$ eV.

The track structure for slow recoil ions is different from the core and penumbra of heavy-ion track structure discussed above. For recoil ions, $r_B$ becomes less than the interatomic distance $a$, in which case $a$ is taken for $a_0$. We assume most δ-rays produced by recoil ions do not have sufficient energy to effectively escape the core and form an undifferentiated core. Then, the radial distribution of the track may be approximated as a single Gaussian and $LET_c$ in Eq. (11) is replaced by $LET_{el}$ for recoil ions. The excitation density can be so high that the number density of excited species $n_i$ estimated can exceed the number density $n_0$ of liquid argon as for fission fragments. When this should occur, redistribution of energy and core expansion may take place, $a_0$ is determined so that $n_i$ does not exceed $n_0$. The initial radial distribution of track core for various ions are shown in Fig. 2. The tack structures due to various ions are also discussed in Ref. 17.

## IV. Biexcitonic quenching

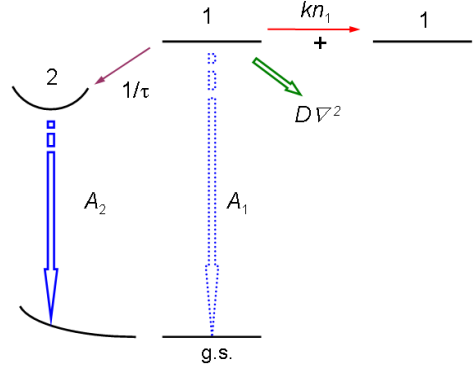

FIG. 3. Schematic diagram illustrating de-excitation and diffusion for the free (suffix 1) and self-trapped (suffix 2) excitons in liquid rare gases. An arrow $kn_1$ shows a biexcitonic collision that is the proposed mechanism for electronic quenching.

In condensed rare gases, both ionization and excitation produced by the ionizing particles, after recombination and/or relaxation, eventually give the lowest $^1\Sigma_u^+$ and $^3\Sigma_u^+$ self-trapped exciton (excimer) states $Ar_2^*$, which scintillate in the vuv region centered at 127 nm through the transition to the repulsive ground state $^1\Sigma_g^+$,

$$Ar_2^* \rightarrow Ar + Ar + h\nu \;. \qquad (13)$$

Since, the lifetimes of the vuv emission do not depend on LET or the existence of quenching [18], quenching occurs prior to self-trapping. Free excitons Ar* and/or highly excited species may be responsible for the quenching. The proposed quenching mechanism is a bi-excitonic collision [10].

$$Ar^* + Ar^* \rightarrow Ar + Ar^+ + e^-(KE) \;. \qquad (14)$$

The ejected electron, e⁻, may immediately lose its kinetic energy $KE$ close to one excitation before recombines with an ion. The overall result is that two excitons are required for one photon. It should be noted that Eq. (14) applies to excitons formed directly or ion recombination.

IV.1. The α-core approximation

The initial radial distributions of excited species in the track core for recoil Ar ions and α particles are similar to each other as shown in Fig. 2. A crude estimate has been made assuming that $q_{el}$ for recoil Ar ions is constant and the same as $q_{\alpha c}$ for α-particles. The approximation was applied before for liquid xenon with considerable success [8] and later for liquid argon [5]. For 5.305 MeV α-particle, taking the fractional energy deposit in the core $T_c/T = 0.72$ and the experimental $q_\alpha = 0.71$ in liquid argon, Eq. (11) is,

$$T_s/T = 0.71 = 0.72 q_{\alpha c} + 0.28 . \qquad (15)$$

This gives $q_{\alpha c} = 0.60$. For recoil Ar ions, $T_s = q_c T_c = q_{el} T$, where $T = E_\eta$. Then we have $q_{el} = q_{\alpha c} = 0.60$. The total quenching factor in the α-core approximation is given as,

$$q_T = q_{el} \cdot q_{nc} = 0.60 \cdot q_{nc} . \qquad (16)$$

The α-core approximation gives $q_T = 0.68 \cdot q_{nc}$ for Xe recoil ions in liquid xenon [8].

IV.2. Diffusion kinetics

The details of the calculation have been described in Ref. 10; therefore, only briefly discussed here. $k$ and $A$ are defined differently in this section from the rest of the paper. The diffusion-kinetic equations for free (index 1) and the self-trapped (index 2) excitons may be expressed as

$$\partial n_1 / \partial t = D\nabla^2 n_1 - k n_1^2 - n_1 / \tau - A_1 n_1 \qquad (17)$$

$$\partial n_2 / \partial t = n_1 / \tau - A_2 n_2 \qquad (18)$$

where $n$ is the exciton density, $D$ is the diffusion coefficient of the free exciton, $k$ is the specific rate of biexcitonic quenching, $\tau$ is the free exciton lifetime against self-trapping. $A_1$ is the radiative decay constant for free excitons and $A_2$ is that for self-trapped excitons. The diffusion-kinetic equations were solved by the method of "prescribed diffusion". In cylindrical geometry one writes with the Gaussian function

$$n_1(r,t) = N(t) G(r,t) \qquad (19)$$

where $N(t)$ is the number of free excitons per unit length at time $t$ and

$$G(r,t) = \frac{\exp(-r^2/a_t^2)}{\pi a_t^2} , \quad a_t^2 = a_0^2 + 4Dt \qquad (20)$$

is a normalized distribution at any time $t$. Eq. (20) provides the spread of the special distribution by diffusion. The initial value for $N(t)$ is given by

$$N(0) = LET_{el} \cdot \rho_l \cdot (1 + N_{ex}/N_i)/W \qquad (21)$$

where $\rho_l$ is the density of the liquid, $N_{ex}/N_i$ is the initial ratio of excitation and ionization, and $W$ is the W-value, an average energy required to produce an electron-ion pair. Since accurate values for slow Ar ions are not available, we took values of $N_{ex}/N_i = 0.21$ [19, 20] and $W = 23.6$ eV [5] reported for electrons in liquid argon.

The excitons are treated as "free" excitons, that particles moving rapidly in condensed argon with a mass similar to the electron mass [21]. The parameters used in the calculation are basically the same as those described before. The rate constant $k$ is given by $k=\sigma v$ where $v$ is the thermal velocity of collision partners ($v = 1.2 \times 10^7$ cm/s) and $\sigma$ is the cross section. The hard-sphere cross section is $\sigma_{HS}$ ~170 Å$^2$. We took $k = k_{HS}/4 = 5 \times 10^{-8}$ cm$^3$/s. The free exciton lifetime $\tau$ is taken to be 1 psec and the diffusion constant $D = 1$ cm$^2$/s is used. The initial radius $a_0 = 3.9$ Å for recoil Ar ions.

We have ignored the radiative term $-A_1 n_1$ in Eq. (17) since the term is orders of magnitude smaller than the self-trapping term $-n_1/\tau$ and the radiative term $-A_2 n_2$ in Eq. (18) since $\tau$ is much shorter than the lifetime of light emission. We calculated the number $N_2(\infty)$ of self-trapped excitons per unit length, which survived quenching. The fraction $N_2(\infty)/N_0$ gives $q_{el}$. We have calculated the number of self-trapped exciton up to $t/\tau=4$.

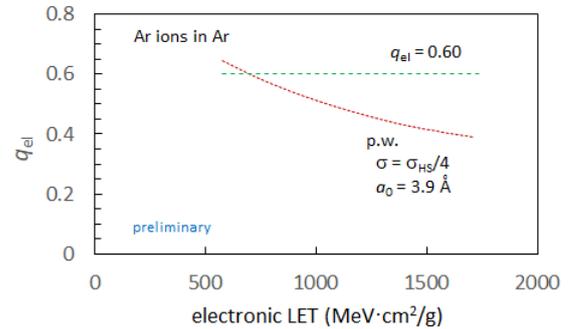

FIG. 4. The electronic quenching factor for slow Ar ions in liquid argon estimated by biexcitonic diffusion kinetics (dotted curve) as a function of electronic LET. The α core approximation (dashed line) is also shown.

Scintillation efficiency for recoil ions as a function of energy is needed for WIMP searches. The electronic quenching factor $q_{el}$ has been calculated and expressed as a function of $LET_{el}$,

$$q_{el} = 0.119 LET_{el}^2 - 0.493 LET_{el} + 0.888 \qquad (22)$$

where $LET_{el}$ is in [MeV·cm$^2$/mg]. This relation is applicable only to slow ions. The result is as shown in Fig. 4.

Then, the total quenching factor at energy $E$ is obtained by integration,

$$q_T = \int_0^E q_{el}(E') \cdot \frac{dE_\eta}{dE'} dE' / \int_0^E dE' \qquad (23)$$

The scintillation yield (or $q_T$) obtained above is defined as the ratio of the nuclear recoil scintillation response to the relativistic heavy ions, which show no reduction due to escaping electrons.

## V. Results

The result obtained for Ar recoil ions in liquid argon is shown in Fig. 5 together with reported experimental results [22-27] as a function of the energy. It should be noted that the experimental values are the scintillation efficiency, RN/γ ratios ($L_{eff}$), relative to 122 keV (or 59.5 keV) γ-rays. $q_T$ is expressed as,

$$q_T = L_{eff} L_{\gamma 0} \quad (24)$$

where $L_{\gamma 0}$ is the scintillation efficiency for γ-rays used as a reference. However, an accurate value for $L_{\gamma 0}$ is not available. The scintillation efficiency for $^{210}$Po 5.305 MeV α-particles $L_\alpha = 0.71$ [28] and the β/α ratio of 1.11 for 1 MeV electrons [28] gives $L_{\gamma 0} = 0.8$. The value is considered to give a lower limit since the LET increases as the energy decreases then $L_{\gamma 0}$ expected to increase. We used $L_{\gamma 0} = 0.83$ by taking into account small differences for $^{22}$Na, $^{133}$Ba and $^{241}$Am γ-rays [27].

The solid curve shows the nuclear quenching $q_{nc}$ taken from Lindhard with $k = 0.15$. The dashed curve is the α-core approximation, $q_T = q_{nc} \cdot q_{el}$, obtained by assuming a constant $q_{el} = 0.60$ as discussed above. The dotted curve is the present calculation. The values for α-core approximation increase with energy, while present calculation shows relatively constant $q_T$ values. $q_{nc}$ increases with energy increase, together with $LET_{el}$ increase, this in turn increase in $N_0$ which make a decrease in $q_{el}$ value. $q_{nc}$ and $q_{el}$ tend to work against to each other, consequently, give relatively constant $q_T$.

The present result (the dotted curve) is smaller than the experimental values; however, reproduces the experimental energy dependence reported by Micro-CLEAN [23] and DARWIN [24] and Creus et al. [26] that show relatively constant scintillation efficiency above ~20 keV. While, the results obtained by SCENE [25] and ARIS [27] were well reproduced by the α core approximation. The measurements, as well as the theory, become extremely difficult at a very low energy region. Experimental results are scattered and have large errors below ~20 keV.

## VI. Discussion

### VI.1. Scintillation yields

The total stopping power $S_T$ for Ar ions in argon increases rapidly with the energy increase up to about

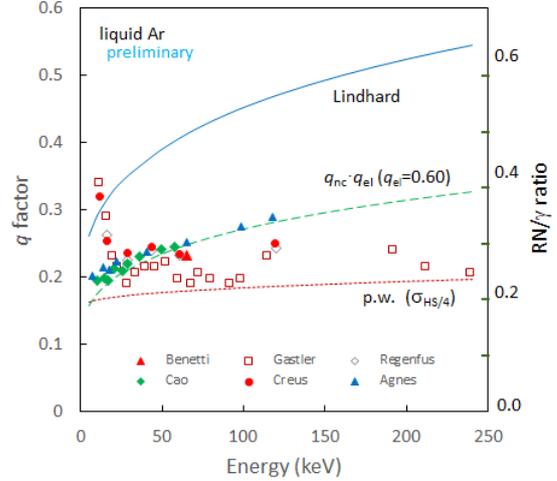

FIG. 5. The quenching factor $q$ and RN/γ ratio for recoil Ar ions in liquid argon as a function of energy. The theoretical values are shown on the left hand axis and the experimental RN/γ ratios [22-27] are shown on the right hand axis, using $L_{\gamma 0} = 0.83$, see the text. The experimental error bars are not shown.

20 keV then becomes almost flat. $S_n$ peaks at around 20 keV then decreases with energy, while $S_e$ contribution increases as $E^{1/2}$ and compensates $S_n$ decrease. Both $S_e$ and $LET_{el}$ increase with energy; however, the magnitude and increasing rate are quite different from each other, particularly at a low energy region. Therefore, it is not appropriate to use $S_e$ in place of $LET_{el}$ as used in most data analysis for recoil ions in liquid rare gases.

The ionization measurements in gas phase may give $q_{nc}$ for rare gases. Number of ion pairs $N_i^g$ produces in gas phase is expressed, to a first approximation

$$N_i^g = E/W \approx q_{nc} E / W_\alpha^g, \quad (25)$$

where $W_\alpha^g$ = 26.4 eV is the $W$-value for α-particles in gaseous argon. Phipps et al. measured $W$-values for 25 - 100 keV Ar ions in argon [29]. The values agreed with Eq. (7) expected by Lindhard theory within ~12 %. A simplification in taking $k = 0.15$, instead of $k = 0.145$, overestimates η values about 3 %. Then the agreement becomes within ~10 %.. Platzman's energy balance equation relates the $W$-value and the $N_{ex}/N_i$ ratio [30],

$$W = \frac{E}{N_i} = \bar{I} + \frac{N_{ex}}{N_i} \bar{E}_{ex} + \bar{E}_{sb} \quad (26)$$

where $\bar{I}$ and $\bar{E}_{ex}$ are the average energy for ionization and excitation, respectively, and $\bar{E}_{sb}$ is the averaged energy spent as the kinetic energy of sub-excitation electrons. The agreement mentioned above supports the assumption that the $N_{ex}/N_i$ ratio for slow ions is the same as that for fast ions (0.21 for liquid argon).

The present calculation gives relatively small $q_T$ values compared with experiments. The main uncertainty in comparing the theory and the experiments is due to $L_{\gamma 0}$ value, 0.83 with +0.12/-0.05. Some measurements assume $L_{\gamma 0} =1$ in analyzing data. The measurement of $S1$ signal at zero fields requires some cautions. The observation time has to be long enough. Even the stray field from PMT influences the scintillation measurements for γ-rays.

The values for the track and collision-reaction parameters used in the calculation were taken from the literature and/or estimated with reasonable consideration [10]. Uncertainties originated from those parameters were discussed in detail in Ref. 10. The quenching cross section σ is the adjustable parameter in present calculation. As mentioned above, $\sigma_{HS}/4$ is obtained by using an averaged $<LET_{el}>$ for various particles [10]. The previous study showed that $\sigma_{HS}/4$ can be regarded as an upper limit. The agreement with the experimental results may be improved by taking a smaller σ that is still reasonable.

The shape of each recoil-ion-track changes because of the scattering. The track is short and detours and may have some branches. The struggling is also large. The present calculation for $q_{el}$ ignored these influences for the simplicity.

VI.2. Field effects

Since the electric field does not influence on the stopping process (consequently not on both $q_{nc}$ and the $N_{ex}/N_i$ ratio), we refer the energy as that given to the electronic excitation $E_\eta$ ($T$ in Chapt. III) in this section unless otherwise mentioned. The scintillation and ionization yields are complementary to each other. An appropriate summation of the scintillation and charge yields gives the energy [28]. Some measurements of the response to incident particles observed $S1$ in the absence of the external field [23, 24] while others measured the sum signal, $S1$ and $S2$, in the presence of the field. The absolute value of charge can be obtained; however, observations for scintillation $S1$ and $S2$ are always associated with uncertainties in the quantum efficiencies and the geometrical factors, etc. $S1$ is usually measured relative to scintillation $S1_0$ in the absence of the field.

The electron thermalization time and the thermalization length are quite large for electron excitation in condensed rare gases. Some fraction of electrons escapes from recombination and do not contribute to scintillation within the observing time scale under no electric field. The scintillation efficiency $L_{\gamma 0}$ for electrons and γ-rays at zero field is given by

$$L_{\gamma 0} = (1 + N_{ex}/N_i - \chi)/(1 + N_{ex}/N_i) \quad (27)$$

where χ is the fraction of escaping electrons at zero field. The quantum efficiency for vuv emission in Eq. (13) is taken to be 1 [31]. In the presence of the electric field **E**, the sum is constant and given as normalized to unity,

$$1 = L_{\gamma 0}\frac{S1}{S1_0} + \frac{Q}{Q_\infty}\frac{1}{1+N_{ex}/N_i}, \quad \mathbf{E} \neq 0 \quad (28)$$

where $S1$ and $S1_0$ are the scintillation yield observed with and without the external field, and $Q$ ($S2$) is the charge collected and $Q_\infty = N_i = E/W$ is the charge produced by an incident particle in unit of electron. The $S1/S1_0 - Q/Q_\infty$ plot gives values for the $N_{ex}/N_i$ ratio and χ.

$$\frac{S1}{S1_0} = \frac{1+N_{ex}/N_i - Q/Q_\infty}{1+N_{ex}/N_i - \chi}, \quad \mathbf{E} \neq 0. \quad (29)$$

Measurements in wide range of the field is necessary to extract the $N_{ex}/N_i$ ratio and χ. Deviation from Eq. (29) at a low field have been reported for 1 MeV electron [28]. Eq. (29) at $Q = 0$ is the invers of $L_{\gamma 0}$, Eq. (27). $L_{\gamma 0}$ is basically given by $S1_0/S1_{int}$, where $S1_{int}$ is the intercept of the $S1$-$S2$ plot on zero $S2$, the $S1$ axis. We obtain $L_{\gamma 0} \sim 0.9$ for $^{85m}$Kr γ-rays (10 keV + 31 keV) in liquid argon from ref. [25]. In liquid xenon, $L_{\gamma 0} = 0.77$ was use in Ref. [8]; however, the $S1$-$S2$ plot reported afterwards gave $L_{\gamma 0} = 0.89$ for 122 keV electrons [32].

At a high LET region, escaping electrons may be negligible. A charge increase is associated with a scintillation decrease. When the absolute number of photons emitted $n_{ph}$ and electrons collected $n_i$ are obtained, the sum $n_{ph} + n_i$ gives the energy and is expressed as [31]

$$n_{ph0} = q_{el0} \cdot (N_i + N_{ex}), \quad \mathbf{E} = 0, \quad (30)$$

$$n_{ph} + n_i = q_{el} \cdot (N_i + N_{ex}), \quad \mathbf{E} \neq 0 \quad (31)$$

where suffix 0 indicates at zero field. Otherwise, the energy is given by $q_{el}T$ with the quenching factor $q_{el}$ expressed as [31]

$$q_{el} = q_{el0}\frac{S1}{S1_0} + \frac{Q}{Q_\infty}\left(\frac{1}{1+N_{ex}/N_i}\right). \quad (32)$$

Then, the energy is obtained by the charge ratio $Q/Q_\infty$ and the scintillation ratio $S1/S1_0$ when $q_{el0}$ is available. A factor $1+N_{ex}/N_i$ appearing in Eq. (31) is due to the fact that the average energy required to produce one photon and one electron-ion pair is 19.6 eV and 23.6 eV for liquid argon, respectively, and are not the same. The measurement obtains $q_T = L_{eff} \cdot L_{\gamma 0} = q_{nc} \cdot q_{el}$ and cannot separate $q_{nc}$ and $q_{el}$. Then, Eq. (32) is modified by multiplying $q_{nc}$ on the both sides of Eq.

(32). A factor $q_{nc}$ appears in front of the 2nd term may be estimated by using Lindhard $q_{nc}$.

No proper theories are available to predict the fraction of electrons collected as a function of the electric field in liquid rare gases. The difficulty arises mainly from the differences in density and drift velocity for positive and negative charge carriers and a lack of information on the spatial distributions in liquid rare gases. The theories of Jaffe [33] and Onsager [34] are inapplicable to electron-ion recombination in liquid rare gases. Phenomenological models, such as Thomas-Imel [35] and Doke-Birks [36,37] are in use often in modified forms, with fitting parameters that have no physical significance. Computer simulation methods [38] are preferable and may be necessary. $LET_{el}$ may give basic information on the initial charge distributions due to recoil ions to start the simulations.

Movements of the energy and charge carriers change drastically at self-trapping [21]. The diffusion constants of the excited states change from ~1 cm/s to $2.43\times10^{-5}$ cm/s for liquid argon. The hole mobility is orders of magnitude higher than that of $Ar_2^+$ in condensed argon. The diffusion length for excitons was reported as 12 nm for solid argon. The initial radial distribution for $Ar_2^*$ and $Ar_2^+$ may be ~10 nm. The effects caused by these may have to be considered in study of the recombination process.

VI.3. General remarks

The measurement in liquid phase can obtain $q_T$; however, it is quite difficult to separate contributions from nuclear quenching and electronic quenching for slow recoil ions. The $W$-value measurements in gas phase should be performed to obtain experimental $q_{nc}$ [39]. The numbers of photons and ion pairs produced are quite limited, the statistics obtained is not enough in most cases. The parameters should be carefully set, otherwise the fitting procedure can lead to an improper behavior of physical quantity and may result in a wrong conclusion. The recombination process in a dense plasma is complex and needs much more detailed investigation.

$LET_{el}$ has wide usage. Scintillation yield for recoil ions in inorganic crystals can be estimated by comparing response at the $LET_{el}$ for slow ions with LET for fast ions [40].

Lindhard has pointed out that the Thomas-Fermi treatment, which the stopping power theories for heavy ion collisions based on, is a crude approximation at very low energy [11]. "The energy loss to electrons is actually correlated to the nuclear collisions, and in close collisions considerable ionization will take place." The distinction between scattering and electronic excitation, Eq. (1), becomes a blur. Full quantum mechanical calculations including the molecular orbit (MO) theory [41] is required. Also, elaborated measurements [42] should be extended to incorporate large scattering angle.

We recommended that terms and nomenclature follow custom in radiation physics and radiation chemistry, otherwise, the knowledge accumulated over decades would be overlooked, apart from confusions wrong usage of terms may cause.

## VII. Summary

The scintillation yields due to Ar ions recoiled by WIMPs in liquid argon has been evaluated. Nuclear quenching was obtained by Lindhard theory. A biexcitonic diffusion-reaction calculation was performed for electronic quenching. A cylindrical track structure was considered. The electronic LET was used to obtain the excitation density needed for the calculation. The present theory gives lower limit for the quenching factor with a quenching cross section of a quarter that the hard-sphere cross section. Most experimental results fall between the present model and a simple α-core approximation within errors. The main uncertainty for comparing the theory and the experiments is due to a lack of accurate scintillation efficiency for low energy γ-rays. The use of the $S1$-$S2$ plot is proposed. Some problems in analyzing experimental data with and without external field were discussed.

## Acknowledgements

We would like to thank Dr. M. Yamashita and Dr. K. Nakamura for useful discussions. The work described herein was supported in part by the Office of Basic Energy Science of the Department of Energy.

Notes.1        The WIMP-Hunter's Guide to Radiation Physics        v.1

**$W$-value**        $W = E/N_i$

The average energy expended to produce an electron-ion pair. Ionizations of all generation are included in the definition of $W$, which is therefore a gross average. It should not be confused with the work function. $W_{ph}$ is defined as the average energy expended to produce one photon.

> the work function is the minimum energy needed to remove an electron from a solid to a point in the vacuum outside the solid surface. a property of the surface of the material.        [Wikipedia]

**$N_{ex}/N_i$**        ($N_{ex}$ and $N_i$ are the initial number of excitons and e-ions pairs, respectively, produced by incident particle.)

can depend on particles and energy. However, $N_{ex}/N_i = 1$ for recoil ions sounds too large. Bookkeeping, like Eqs. (26)–(32) in present paper, should be done properly. Some refers ref. [36] for $N_{ex}/N_i = 1$, however, the paper reads "the ratio of $aS_0$ to $N_{ex} + N_i$ should be unity". While, "the mean energy for an ionization is approximately equal to the mean energy for an ionization, which means that $N_{ex}/N_i \approx 1$." No, it doesn't mean that. The ratio is determined by the oscillator strength [A1]. It implies however, together with the triplet state is not 'real triplet' in Ar and Xe, the $N_{ex}/N_i$ ratio does not depend strongly on the kind of particles and the energy.

**Stopping power and LET** (linear energy transfer)

When you say stopping power, you are on the incident particle, you don't care what secondary particles you kicked out do afterwards. When you say LET, you are in the target material, gather every secondary particle. They are the same for fast ions.

**Birks's Law**

$$\frac{dL}{dx} = \frac{A dE/dx}{1 + B dE/dx} \quad \Rightarrow \quad \frac{dL}{dE} = \frac{1}{1 + B dE/dx}$$

an empirical formula for the scintillation yield per path length as a function of the stopping power for organic scintillators, which do not have escaping electrons. it gives large scintillation yield at low LET and small yield at high LET. The scintillation mechanism is complicated in organic scintillator therefore an empirical formula is useful. Formula on the right is obtained that in the limiting cases, $dL/dE = 1$ for $dE/dx \to 0$ and $dL/dE = 0$ for $dE/dx \to \infty$. Often, the averaged values $<dL/dE>$ and $<dE/dx>$ are used in place of $dL/dE$ and $dE/dx$, respectively. It is applicable to inorganic scintillators in the high LET region. If you ever want to use Birk's law for recoil ions, $dE/dx$ should be replaced by $LET_{el}$ since quenching is electronic process. The electronic stopping power $(dE/dx)_e$ does not give the number of excited species (including ions) par path length, see Fig. 1.

**Thermalization time $\tau_{th}$ and thermalization length $R_{th}$.**

The electron has to lose its kinetic energy before recombination. The elastic scattering brings the energy down to ~ 1 eV quickly. Below ~ 1 eV the elastic scattering becomes inefficient. Condensed Ar and Xe do not have efficient energy-loss mechanisms such as through molecular vibration and rotation. As a result, $\tau_{th}$ in liquid Ar and Xe are exceptionally long, ~ ns [A2]. The scintillation time profile shows this in the rise time for electrons [18]. During the thermalization process, the ionized electron can travel far from the parent ion. The thermalization distance $R_{th}$ becomes longer than the Onsager radius and the average ion distance. The saturation curve, the fraction of charge collected as a function of the electric field, for 1 MeV electron, is different from that for isolated ion pairs [A3]. The Onsager theory is inapplicable to even for minimum ionizing particles in LAr and LXe.

However, the rise time due to α-particles and fission fragments is quick [18]; the electron recombines much faster than $\tau_{th}$. The recombination and charge collection have to be treated separately for recoil ions and γ-rays.

**Recombination theory**
gives the saturation curve, the fraction of charge collected $Q/Q_\infty$, where $Q_\infty=N_i$, ($e$=1) as a function of the electric field. No proper theories are available for condensed rare gases. The data should be collected such that $Q$ comes near to saturate to obtain useful values for parameters and $Q_\infty$ when a fitting function is used. Required electric field $F$ for γ-rays may be ~12 kV/cm and ~6 kV/cm or more for LAr and LXe, respectively. The condition never achieved for α-particles and recoil ions, data obtained for γ-rays may be used for the charge calibration. A small ionization chamber with windows for PMT may be used.

**Thomas-Imel Box model**
This phenomenological model has originally been applied to ~1 MeV electrons since Onsager's geminate theory is inapplicable to LAr and LXe. The box model assumes that the e-ion pairs are isolated and uniformly populated in a box, and the positive ion mobility is orders of magnitude smaller than the electron mobility. I-T does not fit for alphas. Later, a double charge-density model consists of, the minimum ionization part and the high-density part near the end of δ-track, is proposed to explain the fact that the measured energy resolution is much larger than that expected from Poisson statics. T-I gives a convenient fitting form.

$$\frac{Q}{Q_0} = \frac{1}{\xi}\ln(1+\xi)$$

**Penning ionization**
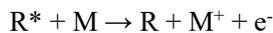
When the excitation potential of R* is higher than the ionization potential of M, R* can ionize M by collision. Penning ionization occurs in mixture. The measurement of Penning ionization Ar* + Xe → Ar + Xe$^+$ + e$^-$ in liquid Ar-Xe mixture gave $N_{ex}/N_i$ = 0.21 for LAr [19].
**Hornbeck-Molnar process** occurs in relatively-high-pressure gas.
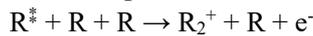
where R$^{\ddagger}$ is higher excited states.

**Singlet and Triplet**
The lowest excited states, $^1\Sigma_u^+$ and $^3\Sigma_u^+$, decay radiatively to the repulsive ground state $^1\Sigma_g^+$ in Ar and Xe. The transition from $^3\Sigma_u^+$ to $^1\Sigma_g^+$ is forbidden by the selection rule $\Delta S = 0$. However, the Russell-Saunders notation is not an adequate representation for heavy atoms. The $^3\Sigma_u^+$ state is not 'true' triplet and not a metastable state. Therefore, behavior of the singlet to triplet ratio S/T for electron and α-particle excitation is different from that in organic compounds which have metastable states.

**The stopping power theory**
Many theories have been proposed to refine or replace the Lindhar theory for slow ion stopping powers. Some uses more realistic potential such as Molière and Lenz-Jensen instead of the Thomas-Fermi potential. However, "the scattering is only quasi-elastic and cannot in detail be described by a potential between two heavy centres." Also, the ion and atom do not come close to each other in those potentials at very low energy."… at extremely low ε-values, $\varepsilon < 10^{-2}$, the nuclear scattering and stopping becomes somewhat uncertain [11]." In this aspect, those theories are no better than the Lindhard theory at very low energy.

**BK model** [A4]
takes a dielectric-response approximation as the LS theory [14]. BK consider an electron gas that is suited for metal and discus the effective charge of projectile. The stopping power for heavy ions uses a scaling rule, $S = (\zeta Z_1)^2 S_p$, where $\zeta Z_1$ is the effective charge and $S_p$ is the stopping power for protons.

**Effective charge**
is determined by the velocities of incident ion and the valence electrons in medium. The effective charge in the Thomas-Fermi model is included in the LS theory as discussed in Ref. [17] and references therein.

**The nuclear quenching factor** $q_{nc}$ (the Lindhard factor)

The stopping powers contain only a part of the necessary information to obtain $q_{nc} = \eta/\varepsilon$. Lindard et al. [7] solved the integral equation and gave numerical results for the nuclear quenching factors. For most cases $k$ is 0.1-0.2, e.g., 0.139, 0.145 and 0.166, respectively, for Ne-Ne, Ar-Ar and Xe-Xe. It does not mean their calculation have such a large uncertainty. The value given by $k = 0.110$ [Hitachi] as often quoted is the total quenching factor $q_T = q_{nc} \cdot q_{el}$. $q_T = q_{nc} \cdot 0.68$ gives values close to $q_{nc}$ for $k = 0.110$ and $q_{nc}$ and $q_{el}$ are difficult to separate [A5].

A relation, $\dfrac{S_e}{S_n + S_e}$ does not give $q_{nc}$.

The difference is obvious when compare the values with the Lindhard values [2] using Fig. 1, they are far off. A modification $aS_e/S_n$, where $a$ is a constant, makes even worse. The integral equations have to be solved to obtain $q_{nc}$ when stopping theories are introduced.
The $W$-value measurements in gas phase give most reliable information on $q_{nc}$.

**Data analysis**

The scintillation yield as a function of energy and the charge yield as a function of the field are usually a simple function. Models with a few free parameters can fit the results. However, that does not necessarily mean the model or the theory is correct. The measured value should be obtained by model-independent way as much as possible. After the values are obtained, a theory or model may be used to interpolate values and/or to compare that results with 'theory'. "XXX" theory or "YYY" model has assumptions and a region where the theory can be applied. That should be respected. Some constant never be a "free" parameter. A careless modification can end up with a fitting function merely looks like "XXX" theory or "YYY" model.

---

## NOMENCLATURE

$\alpha$ is customary used for the recombination coefficient
$\varepsilon, \eta, \upsilon$: the dimensionless energies
$\varepsilon$: dielectric constant

quenching
refers to process which decreases scintillation yield. It is recommended to use 'quenching' in high LET in condensed rare gases. The decrease in scintillation yield at low LET is due to escape electrons which is completely different from process at high LET.

$N_{ex}/N_i$ , $W$, $W_{ph}$, as mentioned above.

---

---

to be refined
to be continued